\documentstyle[11pt,paspconf,epsf]{article}

\begin{document}

\title{Investigating the process of star formation in young LMC star clusters}
\author{R.A. Johnson, S.F. Beaulieu, R.A.W. Elson, G. Gilmore}
\affil{Institute of Astronomy, Madingley Road, Cambridge, UK}
\author{N. Tanvir}
\affil{University of Hertfordshire, College Lane, Hatfield, UK}
\author{B. Santiago}
\affil{Universidade Federal do Rio Grande do Sul, Porto Alegre, Brazil}

\begin{abstract}
The rich star clusters in the Large Magellanic Cloud (LMC) are ideal
for studying the process of star formation.  Here we focus on the
determination of age spreads amongst the massive stars in two young
clusters, NGC1818 and NGC1805. We present colour magnitude diagrams
(derived from HST data) for these clusters, and discuss the
difficulties in age spread determination.

\vspace*{0.2cm}
\noindent Project Web Page - www.ast.cam.ac.uk/LMC
\end{abstract}

\section{Introduction}
This work is part of a large cycle 7 HST project (PID 7307) to study
the formation and evolution of rich star clusters in the LMC. Details
and first results can be found in Beaulieu (1998),
Elson (1998c) and Johnson (1998).

We have obtained observations (with WFPC2, NICMOS and
STIS) of 8 rich LMC star clusters. These clusters have masses
$\approx$ 10$^{4}$M$_{\sun}$\ and ages from 20 Myr to 12 Gyr. They are
grouped into four age pairs and we concentrate here on the
youngest cluster pair of NGC1818 and NGC1805. NGC1818 is quite well
studied (Will 1995, Hunter 1997, Elson 1998a, Elson 1998b) and has a
metallicity [Fe/H] $\approx$ -0.8, a dynamical time of 1.6-10 Myr and
a relaxation time of 150-600 Myr (Elson 1987). Little previous work exists
for NGC1805.

For the young clusters the main aims of the project are: to search for
primordial binaries (Elson 1998a, Santiago 1999) and pre-main sequence
stars (Beaulieu 1999), and to place limits on the age spreads
amongst the massive stars (Johnson 1999). From the low mass and
high mass star formation timescales that we find we will 
investigate the sequence of star formation.

In this paper we focus on searching for age spreads amongst the
massive stars. The amount of age spread depends on the timescale for
star formation which has important implications for the mechanism
which triggered the star formation and for the early evolution of the
cluster. A short (compared to the dynamical time) timescale requires a
strong perturbation to initiate it, whereas self propagating star
formation proceeds on the order of the dynamical time. The combination
of the star formation timescale and the efficiency of the star
formation influences the cluster evolution and possible disruption.
The various scenarios are discussed in Elson (1987).
The amount of age spread also has implications for the process of self
enrichment in these clusters which may be expected to occur on the
order of a massive star lifetime, $\approx$10 Myr.

\section{Results}
To derive the amount of age spread amongst the massive stars we use
colour magnitude diagrams produced from the WFPC2 data.  Isochrones
with the same metallicity but different ages diverge for the massive
stars (V$<$18.5) producing a colour spread. For example, at V=16,
stars that are 25 and 40 Myr old differ by $\approx$0.04 mags in V-I
which is similar to our photometric errors.

The WFPC2 images of these clusters (see the project web page) were
oriented with the cluster core (and hence the majority of the massive
stars) falling on the planetary camera (PC) chip.  Here we just
present results from the PC, results from all the chips will be
presented in Johnson (1999).

\begin{figure}
\plotfiddle{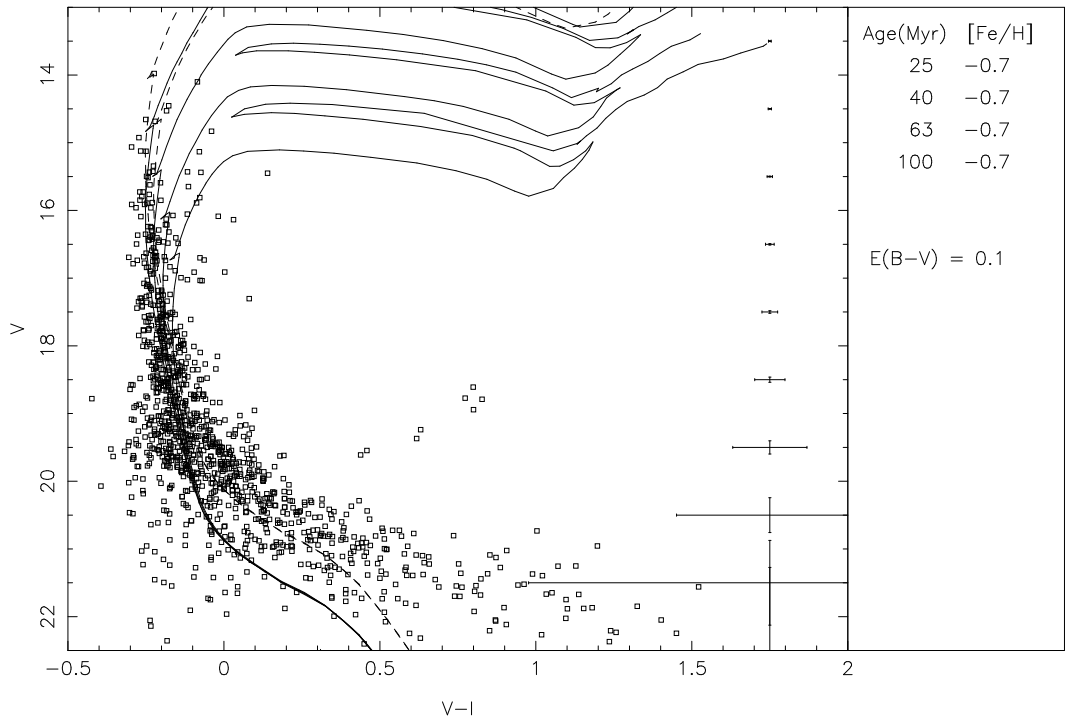}{8.5cm}{-90}{54}{54}{-240}{330}
\plotfiddle{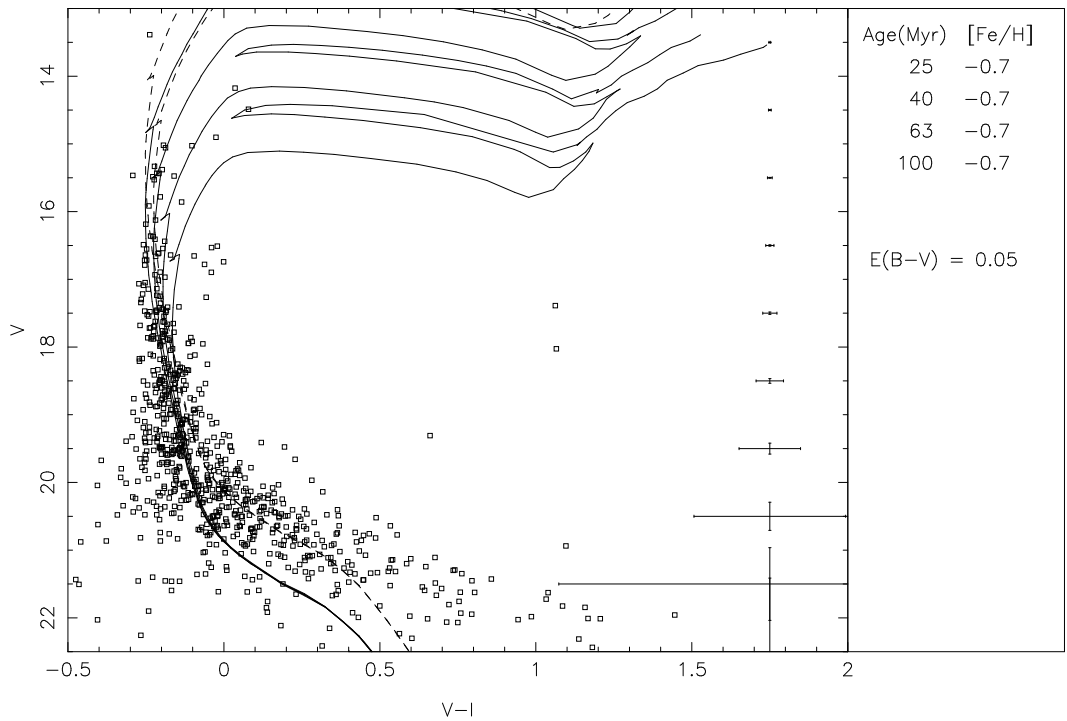}{8.5cm}{-90}{54}{54}{-240}{305}
\caption{V vs V-I colour magnitude diagrams for the PC data of the centre of NGC1818 (top) and NGC1805(bottom)}
\label{fig:colmag}
\end{figure}

Figure~\ref{fig:colmag} shows V vs V-I colour magnitude diagrams for
NGC1818 and NGC1805. 
Overlaid on the data (solid lines) are four isochrones (Bertelli 1994), all
with [Fe/H]=-0.7, and ages of 25, 40, 63 and 100 Myr. NGC1818 is known
to contain binaries (Elson 1998a). The dashed lines show the position
of an equal mass binary sequence for the 25 and 40 Myr isochrones.

It can be seen from these figures that the massive stars do exhibit a
large colour spread. However, before attributing this to age, other
possibilities have to be ruled out. Possible contributors to the
colour spread are binaries, differential stellar rotation and Be
stars.  Further complications are provided by interacting binaries
which make young stellar populations appear younger (in colour
magnitude diagrams) than they really are (e.g. Van Bever 1998).

\begin{figure}
\plotfiddle{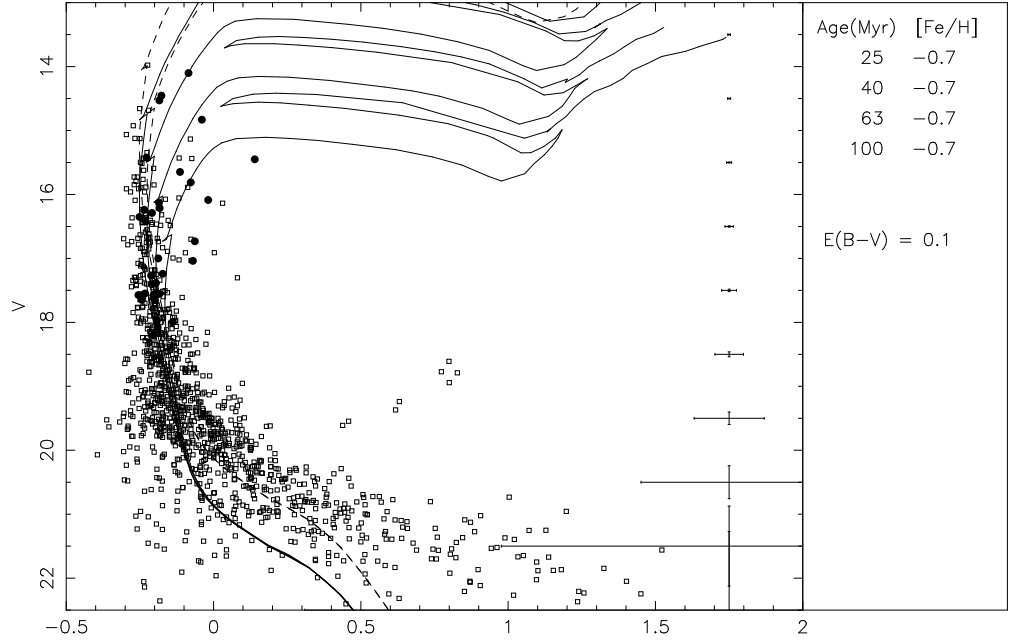}{7.7cm}{-90}{51}{51}{-230}{280}
\caption{NGC1818 V vs V-I colour magnitude diagrams with Be stars identified
by Grebel (1997) marked with filled circles}
\label{fig:bestars}
\end{figure}

We aim to simulate colour magnitude diagrams including these effects
to try and disentangle the various possibilities (Johnson 1999).
However, it is clear from previous observations (Grebel 1997) that Be
stars make a large contribution to the colour spread.  Be stars are
stars that have at some time shown H$\alpha$ emission.  It is thought
that this emission comes from a circumstellar disk which also reddens
the star colours (e.g. Fabregat et al. 1996).
Figure~\ref{fig:bestars} shows the positions on the NGC1818 colour
magnitude diagram of the Be stars identified by Grebel (1997). It can
be seen that these stars show a wide spread in colour, from normal B
star colours to some 0.5 mags redder in V-I.  There are still several
very red stars in the colour magnitude diagram that are not identified
as Be stars by Grebel.  Unfortunately the Be star phenomenon is known
to vary on short timescales, therefore some stars that are Be stars
(and therefore redder) at the time of our HST observations may have
appeared as normal B stars at the time of Grebel's observations. To
eliminate this problem we have obtained time on the ESO NTT to make
our own Be star identifications.

\begin{references}
\reference Beaulieu, S.F., Elson, R.A.W., Gilmore, G., Johnson, R.A., Tanvir, N., Santiago, B. 1999, \mnras, in prep.
\reference Bertelli, G., Bressan, A., Chiosi, C., Fagotto, F., Nasi, E. 1994, \aaps, 106, 275
\reference Elson, R., Fall, S.M., Freeman, K. 1987, \apj, 323, 54
\reference Elson, R.A.W., Sigurdsson, S., Davies, M., Hurley, J., Gilmore, G. 1998a, \mnras, 300, 857
\reference Elson, R.A.W., Sigurdsson, S., Hurley, J., Davies, M.B., Gilmore, G.F. 1998b, \apjl, 499, L53
\reference Elson, R., Tanvir, N., Gilmore, G., Johnson, R.A., Beaulieu, S. 1998c, IAU Symp. 190, 40
\reference Fabregat, J., Torrej\'{o}n, J.M., Reig, P., Bernabeu, G., Busquets, J., Marco, A., Reglero, V. 1996, \aaps, 119, 271
\reference Grebel, E.K. 1997, \aap, 317, 448
\reference Hunter, D.A., Light, R.M., Holtzmann, J.A., Lynds, R., O'Neil Jr., E.J., Grillmair, C. 1997, \apj, 478, 124
\reference Johnson, R.A., Beaulieu, S.F., Elson, R.A.W., Gilmore, G., Tanvir, N., Santiago, B. 1999, \mnras, in prep.
\reference Santiago, B., Beaulieu, S.F., Elson, R.A.W., Gilmore, G., Johnson, R.A., Tanvir, N. 1999, \mnras, in prep. 
\reference Van Bever, J. \& Vanbeveren, D. 1998, \aap, 334, 21
\reference Will, J.M., Bomans, D.J., de Boer, K.S. 1995, \aap, 295, 54
\end{references}
\end{document}